\documentclass[12pt]{iopart}
\usepackage{amssymb}
\usepackage{txfonts}
\usepackage{iopams}
\usepackage{graphicx}

\begin{document}

\title{Stronger uncertainty relations with improvable upper and lower bounds}
\author{Jun Zhang$^1$, Yang Zhang$^1$, Chang-shui Yu$^1$$^a$ \\[1ex]
  \footnotesize
  \begin{minipage}{9cm}
  $^1$School of Physics and Optoelectronic Technology, \\
  Dalian University of Technology, Dalian 116024, China\\
    \texttt{E-mail:quaninformation@sina.com; ycs@dlut.edu.cn}\\[1ex]
    \\
  \texttt{}
    \end{minipage}
}
\begin{abstract}
We utilize quantum superposition principle to establish the improvable upper and lower bounds on the stronger uncertainty relation, i.e., the  "weighted-like"  sum of the variances of observables.  Our bounds include some free parameters which not only guarantee the nontrivial bounds but also can effectively control the bounds as tightly as one expects. Especially, these parameters don't obviously depend on the state and observables. It also implies one advantage of our method that any nontrivial bound can always be more improvable. In addition, we generalize both bounds to the uncertainty relation with multiple observables, but the perfect tightness is not changed. Examples are given to illustrate the improvability of our bounds in each case.
\end{abstract}

\maketitle
\section{Introduction}

Uncertainty relation (principle) as one of the well-known fundamental principles in quantum mechanics shows that anyone is not able to simultaneously specify the values of the non-commuting and canonically conjugated observables \cite{HUP} . It is also one of the keys that highlights the difference between the quantum and classical worlds. Uncertainty relation not only has a wide range of applications such as entanglement detections \cite{HFH,OG} , security analysis \cite{sa,Wehner} , signal processing \cite{signal,huang} and so on, but also has been developed based on different considerations such as the noise-disturbance uncertainty \cite{Ozawa1,Ozawa6,Ozawa7,Ozawa9} , the entropic uncertainty \cite{Wehner,Deutsch,Kraus,Uffink,Renyi,inade,k-entropy,smooth,fan,shu,qiao}, the uncertainty with memory \cite{memory,ZY} , the uncertainty of successive measurements \cite{SPMq,SPM,ZYQIP} , the uncertainty relation based on Wigner-Yanase skew information \cite{Luo,Luo1,Luo2,Skew} or joint measurements \cite{Werner2,Werner5,Werner6,Wernerz} and so on \cite{Fan,Fan1,Fan2}. In
fact, uncertainty relations and uncertainty principles have been well distinguished in Ref. \cite{peres} . The former refers to the measurements performed on the identical preparations of the state, so one measurement can never induce the disturbance on the other measurement; The latter considers the measurement disturbance induced by the apparatus or the impossible joint measurements of incompatible observables \cite{pati} . Here we mainly consider the uncertainty relations. The most popular version of the uncertainty relation (in many textbooks) is the Kennard-Robertson inequality \cite{kennard,RHUP} which shows the product of the variances of two observables are lower bounded as follows.
\begin{equation}
\Delta X\Delta Y\geqslant \frac{1}{2}\left\vert \left\langle \Psi
\right\vert [X,Y]\left\vert \Psi \right\rangle \right\vert ,  \label{1a}
\end{equation}
where $\Delta X=\sqrt{|\langle \Psi|(X-\langle X \rangle)^{2}|\Psi\rangle|}$ represents the standard deviation of the observable $X$ and $[X,Y]=XY-YX $ stands for the commutator. It is obvious that the right-hand-side vanishes if the expectation value of the commutator vanishes, even though $X$ and $Y$ don't commute with each other. Later the inequality (\ref{1a}) was further improved by Schr$\ddot{o}$dinger who added an anti-commutator contribution into the right-hand-side of Eq. (\ref{1a}). That is,
\begin{equation}
\Delta X\Delta Y\geqslant \frac{1}{2}\left\vert \left\langle \Psi
\right\vert [X,Y]\left\vert \Psi \right\rangle \right\vert+\left\vert\frac{1%
}{2}\{X,Y\}_+-\left\langle X\right\rangle\left\langle
Y\right\rangle\right\vert ,  \label{2a}
\end{equation}
where $\{\cdot,\cdot\}_+$ denotes the anti-commutator. However, it can be easily found that both Eq. (\ref{1a}) and Eq. (\ref{2a}) will provide a trivial lower bound if the state $\left\vert\Psi\right\rangle$ is one eigenstate of $X$ or $Y$, even though $X$ and $Y$ have no common eigenstate.

Recently, L. Maccone and A. K. Pati \cite{pati} have presented a stronger uncertainty inequality which provides a "nontrivial" lower bound on the sum of variances of two observables instead of their product. This uncertainty relation can be formulated as \cite{pati,pati2}
\begin{equation}
\Delta X^{2}+\Delta Y^{2}\geqslant \max \{\mathcal{L}_{1},\mathcal{L}_{2}\},
\label{patis}
\end{equation}%
with
\begin{eqnarray}
\mathcal{L}_{1} &=&\pm i\langle \lbrack X,Y]\rangle +|\langle \psi |X\pm
iY|\psi ^{\perp }\rangle |^{2},  \label{pati1} \\
\mathcal{L}_{2} &=&\frac{1}{2}\Delta (X+Y)^{2},  \label{pati2}
\end{eqnarray}%
where the state $|\psi ^{\perp }\rangle $ is orthogonal to the state $|\psi\rangle $ and the $\pm $ in Eq. (\ref{pati1}) is selected such that the term is positive. When more than two observables are considered, the relevant uncertainty relations are also introduced \cite{SRfei,SRfeiw} . Since all the aforementioned uncertainty relations (Eqs. (\ref{1a},\ref{2a},\ref{patis})) are just the quantitative expression for the nonexistence of common eigenstates in incompatible observables, one can find that the lower bound (Eq. (\ref{patis}))  in qubit system is just a rewriting of the left-hand side of Eq.(\ref{patis}) for a qubit system, even though it could also be of significance in some particular cases. The details about this analysis and the generalized rewriting for all systems are given in appendix. So it remains an interesting question how to present genuine tight enough bounds instead of the potential rewriting.

In this paper, we give a perfect answer to this question by presenting the improvable and nontrivial upper and lower bounds on the "weighted-like" sum of variances of two observables based on the quantum superposition principle. The characteristic of our bounds is some free adjustable parameters included in the bounds, which leads to three obvious advantages: (1) The bounds can always be adjusted to be nontrivial at any case; (2) The improvability of the bounds can be effectively controlled by the free parameters as we expect; (3) All the free parameters do not obviously depend on the state and observables covered in the uncertainty relation. Since our method can introduce free parameters into the bounds, another potential advantage
is that any nontrivial bound can always serve as a tight enough bound. This advantage can be well manifested in the case of multiple observables. As we know, uncertainty relations with two observables can automatically induce the corresponding uncertainty relations with more than two observables, however, one usually needs to consider the different combinations of observables and find their maximum value in order to obtain a improvable bound. This is quite complicated with the increasing number of observables. But it can be found that if employing our method, one can only consider any one simple and nontrivial bound. Here we also provide the alternative upper and lower bounds of the sum of their variances. It is shown that these two bounds have also the controllable tightness. The improvability of our bounds for the uncertainty relation with two or more observables are demonstrated by an example, respectively.

\section{The improvable upper and lower bounds for uncertainty relation with two observables}

To begin with, let's consider two observables $A$ and $B$ operated on a normalized quantum state $\left\vert \psi \right\rangle $. So the variances of them are given by $\Delta A^{2}=\langle \psi |A^{2}|\psi \rangle -\langle\psi |A|\psi \rangle ^{2}$ and $\Delta B^{2}=\langle \psi |B^{2}|\psi\rangle -\langle \psi |B|\psi \rangle ^{2}$. One can construct another pair of unnormalized quantum state $\left\vert \psi _{1}\right\rangle $ and $\left\vert \psi _{2}\right\rangle $ as
\begin{eqnarray}
\left\vert \psi _{1}\right\rangle &=&(A-\left\langle A\right\rangle
)\left\vert \psi \right\rangle ,  \label{s1} \\
\left\vert \psi _{2}\right\rangle &=&(B-\left\langle B\right\rangle
)\left\vert \psi \right\rangle .  \label{s2}
\end{eqnarray}%
It can be easily found that
\begin{eqnarray}
\Delta A^{2} &=&\Vert \left\vert \psi _{1}\right\rangle \Vert ^{2},
\label{nv1} \\
\Delta B^{2} &=&\Vert \left\vert \psi _{2}\right\rangle \Vert ^{2},
\label{nv2}
\end{eqnarray}%
with $\left\Vert \cdot \right\Vert $ representing the $l_{2}$ norm of a state vector. With these notations, we can give our main result in the following rigorous way.

\textbf{Theorem 1}: For a pair of complex number $x=x_{1}+ix_{2}$ and $y=y_{1}+iy_{2}$ with subscripts 1 and 2 representing the real and the imaginary part, respectively (similar notation for other parameters), the stronger uncertainty relation for observables $A$ and $B$ operated on the normalized state $\left\vert \psi \right\rangle $ can be given by
\begin{equation}
\mathcal{B}_{+}\geqslant |x|^{2}\Delta A^{2}+|y|^{2}\Delta B^{2}\geqslant
\mathcal{B}_{-},  \label{main}
\end{equation}%
where $\mathcal{B}_{\pm }$ are the arbitrarily improvable upper and lower bounds, respectively, and given by
\begin{equation}
\mathcal{B}_{\pm }=\frac{1}{2}\left[ \mathcal{B}^{2}(x,y)+\left( |a|\mathcal{%
B}(m,n)\pm |b|\mathcal{B}(\tilde{m},\tilde{n})\right) ^{2}\right] ,
\label{bpm}
\end{equation}%
with
\begin{eqnarray}
\mathcal{B}^{2}(\alpha ,\beta ) &=&\Delta \left( \alpha _{1}A+\beta
_{1}B\right) ^{2}+\Delta \left( \alpha _{2}A+\beta _{2}B\right) ^{2}
+i\left( \alpha _{1}\beta _{2}-\beta _{1}\alpha _{2}\right) \left\langle %
\left[ A,B\right] \right\rangle .  \label{bdef}
\end{eqnarray}%
Here $(\alpha ,\beta )$ represents $(x,y)$ and the free complex number pairs $(m,n)$ and $(\tilde{m},\tilde{n})$ which satisfy
\begin{eqnarray}
x &=&am+b\tilde{m},  \label{lim1} \\
-y &=&an+b\tilde{n},  \label{lim2}
\end{eqnarray}%
and $|ab|\neq 0$.

\textbf{Proof}. At first, one should note that different $x$ and $y$ define different "weight-like" uncertainty relations. Give such an uncertainty relation, $x$ and $y$ should be fixed, but the so-call free parameters throughout the paper don't include $x$ and $y$. We will prove this theorem by three steps. Firstly, we will prove that the sum of the variances are well bounded by $\mathcal{B}_\pm$; Secondly, we will show that $\mathcal{B}_\pm$ are improvable, that is, they can arbitrarily approach the sum of the variances; Thirdly, we will show that Eq. (\ref{main}) will not provide trivial bounds unless $\left\vert\psi\right\rangle$ is the common eigenstate of the observables $A$ and $B$.

(\emph{Bounds}).- Based on Eq. (\ref{s1}) and Eq. (\ref{s2}), one can construct the unnormalized states%
\begin{eqnarray}
\vert \bar{\psi }\rangle &=&x\vert \psi _{1}\rangle+y\vert \psi _{2}\rangle ,  \label{fs1} \\
|\tilde{\psi}\rangle &=&x\vert \psi _{1}\rangle -y\vert \psi_{2}\rangle .  \label{fs2}
\end{eqnarray}%
Alternatively, one can reconstruct $\vert\tilde{\psi}\rangle $ as
\begin{equation}
\vert\tilde{\psi}\rangle =a\vert\tilde{\psi}_{1}\rangle +b\vert\tilde{\psi}_{2}\rangle,  \label{recs}
\end{equation}%
with
\begin{eqnarray}
\vert\tilde{\psi}_{1}\rangle &=&m\vert \psi_{1}\rangle +n\vert \psi _{2}\rangle,  \label{rs1} \\
\vert\tilde{\psi}_{2}\rangle &=&\tilde{m}\vert \psi_{1}\rangle +\tilde{n}\vert \psi _{2}\rangle.  \label{rs2}
\end{eqnarray}%
Substitute Eqs. (\ref{rs1}) and (\ref{rs2}) into Eq. (\ref{recs}) and compare it with Eq. (\ref{fs2}), one will easily find that all the complex parameters $m$, $n$, $\tilde{m}$, $\tilde{n}$, $a$, $b$, $x$ and $y$ are just constrained by Eqs. (\ref{lim1}) and (\ref{lim2}) since Eq. (\ref{recs}) and Eq. (\ref{fs2}) are the same state.

On the basis of the $l_{2}$ norm of a quantum state (similar to Eqs. (\ref{nv1}) and (\ref{nv2})), we have, for $\left\vert \psi \right\rangle $ and $\vert \tilde{\psi}\rangle $,
\begin{eqnarray}
&&\Vert \vert \bar{\psi}\rangle \Vert ^{2}=\left\Vert
x\left\vert \psi _{1}\right\rangle +y\left\vert \psi _{2}\right\rangle
\right\Vert ^{2} =\left\Vert x\left\vert \psi _{1}\right\rangle \right\Vert
^{2}+\left\Vert y\left\vert \psi _{2}\right\rangle \right\Vert ^{2}+x^{\ast
}y\langle \psi _{1}|\psi _{2}\rangle +xy^{\ast }\langle \psi _{2}|\psi
_{1}\rangle ,  \label{21} \\
&&\Vert |\tilde{\psi}\rangle \Vert ^{2}=\left\Vert x\left\vert \psi
_{1}\right\rangle -y\left\vert \psi _{2}\right\rangle \right\Vert ^{2}
=\left\Vert x\left\vert \psi _{1}\right\rangle \right\Vert ^{2}+\left\Vert
y\left\vert \psi _{2}\right\rangle \right\Vert ^{2}-x^{\ast }y\langle \psi
_{1}|\psi _{2}\rangle -xy^{\ast }\langle \psi _{2}|\psi _{1}\rangle .
\label{22}
\end{eqnarray}%
Add Eq. (\ref{22}) to Eq. (\ref{21}), one will arrive at
\begin{equation}
\Vert \vert \bar{\psi}\rangle \Vert ^{2}+\Vert |\tilde{%
\psi}\rangle \Vert ^{2}=2\left\Vert x\left\vert \psi _{1}\right\rangle
\right\Vert ^{2}+2\left\Vert y\left\vert \psi _{2}\right\rangle \right\Vert
^{2}.  \label{preres}
\end{equation}%
In addition, the $l_{2}$ norm of $\vert \bar{\psi}\rangle $ can be expanded as
\begin{eqnarray}
&&\Vert \vert \bar{\psi}\rangle \Vert ^{2}=\left\Vert
x\left\vert \psi _{1}\right\rangle +y\left\vert \psi _{2}\right\rangle
\right\Vert ^{2}  \nonumber \\
&=&\left\Vert \left( x_{1}+ix_{2}\right) \left( A-\left\langle
A\right\rangle \right) \left\vert \psi \right\rangle +\left(
y_{1}+iy_{2}\right) \left( B-\left\langle B\right\rangle \right) \left\vert
\psi \right\rangle \right\Vert ^{2}  \nonumber \\
&=&\Vert \left( x_{1}A+y_{1}B\right) -\left( x_{1}\left\langle
A\right\rangle +y_{1}\left\langle B\right\rangle \right) +i\left[ \left(
x_{2}A+y_{2}B\right) -\left( x_{2}\left\langle A\right\rangle
+y_{2}\left\langle B\right\rangle \right) \right] \left\vert \psi
\right\rangle \Vert ^{2}  \nonumber \\
&=&\Delta \left( x_{1}A+y_{1}B\right) ^{2}+\Delta \left(
x_{2}A+y_{2}B\right) ^{2}+i\left( x_{1}y_{2}-y_{1}x_{2}\right) \left\langle
\left[ A,B\right] \right\rangle  \nonumber \\
&=&\mathcal{B}^{2}\left( x,y\right) .  \label{psie}
\end{eqnarray}
It is obvious that $\mathcal{B}^{2}\left( x,y\right) $ is just given by Eq. (\ref{bdef}). Analogously, the $l_{2}$ norm of $\vert \tilde{\psi}\rangle $ can be rewritten by
\begin{eqnarray}
&&\Vert |\tilde{\psi}\rangle \Vert ^{2}=\Vert a|\tilde{\psi}_{1}\rangle +b|%
\tilde{\psi}_{2}\rangle \Vert ^{2} =\Vert a|\tilde{\psi}_{1}\rangle \Vert
^{2}+\Vert b|\tilde{\psi}_{2}\rangle \Vert ^{2}+2\mathrm{Re}\left( a^{\ast
}b\langle \tilde{\psi}_{1}|\tilde{\psi}_{2}\rangle \right) .
\end{eqnarray}%
According to the Cauchy-Schwartz inequality, $2\mathrm{Re}(a^{\ast }b\langle\tilde{\psi}_{1}|\tilde{\psi}_{2}\rangle )$ can be well bounded as
\begin{equation}
-2\vert a\vert \vert b\vert \Vert |\tilde{\psi}\rangle%
_{1}\Vert \Vert |\tilde{\psi}_{2}\rangle\Vert \leq 2\mathrm{Re}%
(a^{\ast }b\langle \tilde{\psi}_{1}|\tilde{\psi}_{2}\rangle )\leq
2\vert a\vert \vert b\vert \Vert |\tilde{\psi}%
_{1}\rangle\Vert\Vert |\tilde{\psi}_{2}\rangle\Vert .
\end{equation}%
So
\begin{equation}
\left( \Vert a|\tilde{\psi}_{1}\rangle \Vert -\Vert b|\tilde{\psi}%
_{2}\rangle \Vert \right) ^{2}\leq \Vert |\tilde{\psi}\rangle \Vert ^{2}\leq
\left( \Vert a|\tilde{\psi}_{1}\rangle \Vert +\Vert b|\tilde{\psi}%
_{2}\rangle \Vert \right) ^{2}.  \label{psitb}
\end{equation}
Considering the states in Eqs. (\ref{rs1}), (\ref{rs2}) and following the same method as Eq. (\ref{psie}), we can expand $\vert \tilde{\psi}_{1}\rangle $ and $\vert \tilde{\psi}_{2}\rangle $ associated with the complex parameters $a$ and $b$ as
\begin{eqnarray}
\Vert a\vert \tilde{\psi}_{1}\rangle \Vert &=&\left\vert a\right\vert \mathcal{B}(m,n), \\
\Vert b\vert \tilde{\psi}_{2}\rangle \Vert &=&\left\vert b\right\vert \mathcal{B}(\tilde{m},\tilde{n}),
\end{eqnarray}%
where $\mathcal{B}(m,n)$ is again the same as Eq. (\ref{bdef}). Substitute Eqs. (\ref{psie}) and (\ref{psitb}) into Eq. (\ref{preres}), one will directly obtain the result given by Eq. (\ref{main}). Next, we first prove that $\mathcal{B}_{\pm }$ are improvable, that is, they can arbitrarily approach the sum of the variances; then we show that Eq. (\ref{main}) won't provide trivial bounds unless $\left\vert \psi \right\rangle $ is the common eigenstate of the observables $A$ and $B$.

\emph{Arbitrarily improvable}.- Through the proof, one can learn that the inequalities in Eq. (\ref{main}) come from the Cauchy-Schwartz inequality on $2\mathrm{Re}(a^{\ast }b\langle \tilde{\psi}_{1}|\tilde{\psi}_{2}\rangle )$. It can be easily found that both the inequalities in Eq. (\ref{main}) will be saturated once $a=0$ or $b=0$, which is especially independent of the parameters $x$, $y$, the observables $A$, $B$ and the state $\left\vert \psi \right\rangle $. In this case, Eq. (\ref{main}) is
just a trivial equality and it does not denote any uncertainty relation. So neither $a=0$ nor $b=0$ is valid in our theorem. However, the trivial case tells us very useful information. Since $\mathcal{B}_{\pm }$ are continuous functions on $\left\vert a\right\vert $ and $\left\vert b\right\vert $ which can be seen from Eq. (\ref{bpm}) and Eq. (\ref{bdef}), $\mathcal{B}_{\pm }$
will infinitely approach $|x|^{2}\Delta A^{2}+|y|^{2}\Delta B^{2}$ if we let $\left\vert a\right\vert $ or $\left\vert b\right\vert $ infinitely approach $0$ (the other free parameters are limited by Eq. (\ref{lim1}) and Eq. (\ref{lim2}).). In this sense, we say that our upper and lower bounds are arbitrarily improvable. The intuitive illustration of the improvability can also be given by an example in FIG. 1.

\emph{Nontrivial}.- To show the bounds are not trivial, one can analyze the bounds $\mathcal{B}_{\pm }$. However, here we will do it by an easy and direct method. At first, we should note that for a given uncertainty relation, the parameters $x$, $y$, the observables $A$ and $B$ and the quantum state $\left\vert \psi \right\rangle $ are fixed. So one will easily show that $|x|^{2}\Delta A^{2}+|y|^{2}\Delta B^{2}$ does not vanish for nonsingular $x$ and $y$ unless $\left\vert \psi \right\rangle $ happens to be the common eigenstate of $A$ and $B$. This is a normal case since $\left\vert \psi \right\rangle $ has been the common state. Now, we suppose $|x|^{2}\Delta A^{2}+|y|^{2}\Delta B^{2}=\epsilon \neq 0$. Based on the perfect tightness shown above, no matter how small $\epsilon $ is, one can always find proper free parameters such that $\mathcal{B}_{+}=\epsilon _{+}>\epsilon >\epsilon _{-}=\mathcal{B}_{-}>0$ and especially we can let $\epsilon _{\pm }$ infinitely approach $\epsilon $. This means that we can always find such a $\epsilon _{-}\neq 0$ so long as $\epsilon \neq 0$, that is, $\epsilon _{-}$ is not trivial. But although the upper bound $\mathcal{B}_{+}$ does not provide an uncertainty relation in general, it serves as a
complementary constraint on the $|x|^{2}\Delta A^{2}+|y|^{2}\Delta B^{2}$. The proof is completed.\hfill{}$\blacksquare$

\begin{figure}[tbp]
\centering
\includegraphics[width=0.5\columnwidth]{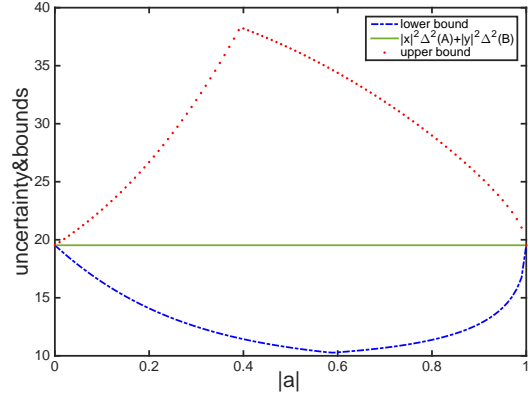}
\caption{(Color online.) The uncertainties of two observable A and B and the
upper and lower bounds vs. the control variable $\left\vert a\right\vert$.
The bounds are always non-negative and the two endpoints show our
arbitrarily improvable upper and lower bounds.}
\end{figure}

To intuitively illustrate the improvability of the bounds, we use Matlab 2014b to randomly generate a 4-dimensional quantum pure state $\left\vert\psi\right\rangle=[0.1452 + 0.3194i,  0.4672 + 0.3066i,  0.3373 +0.5010i,  0.2174 + 0.3905i]^T$ and two observables $A=A_1+A_2i$ and $B=-B_1+B_2i$ with
\begin{eqnarray}
A_1 &=&\left(
\begin{array}{cccc}
5.2528 & 4.1553 & 1.2229 & 3.0871 \\
4.1553 & 5.0443 & 1.1295 & 3.0669 \\
1.2229 & 1.1295 & 0.8441 & 1.2898 \\
3.0871 & 3.0669 & 1.2898 & 3.6033%
\end{array}%
\right) ,
\end{eqnarray}
\begin{eqnarray}
A_2&=&\left(
\begin{array}{cccc}
0 & - 1.6562 & - 0.2396 & - 0.3176 \\
1.6562 & 0 & - 0.1069 & 0.1638 \\
0.2396 & 0.1069 & 0 & 0.3284 \\
0.3176 & - 0.1638 & - 0.3284 & 0%
\end{array}%
\right) ,
\end{eqnarray}
\begin{eqnarray}
B_1 &=&\left(
\begin{array}{cccc}
0.9238 & 1.0856 & 0.6217 & 0.3696 \\
1.0856 & 2.1550 & 1.1369 & 0.5446 \\
0.6217 & 1.1369 & 0.6471 & 0.2780 \\
0.3696 & 0.5446 & 0.2780 & 0.1765%
\end{array}%
\right) ,
\end{eqnarray}
and
\begin{eqnarray}
B_2&=&\left(
\begin{array}{cccc}
0 & - 1.0209 & - 0.0365 & 0.8770 \\
1.0209 & 0 & 1.0103 & 1.0176 \\
0.0365 & - 1.0103 & 0 & 0.3580 \\
- 0.8770 & - 1.0176 & - 0.3580 & 0%
\end{array}%
\right) .
\end{eqnarray}
In addition, we also randomly generate two complex numbers $x=0.267+0.769i$ and $y=-0.234+0.158i$ (throughout the paper all the randomly generated numbers come from Matlab 2014b. We won't mention it again). In order to show that the two bounds can reach the exact value of the sum of variances when $a$ or $b$ vanishes, we impose another condition, $|a|^2+|b|^2=1$. All the other free parameters are randomly generated. Here, we calculate the bounds in two cases. In the first case, we let these parameters be separately given as $m=1.231-0.317i$, $n=1.920+0.701i$, and $\tilde{m}=\tilde{n}=0.1$ when $a=0$. In the second case, we let the parameters be given as $\tilde{m}=0.501+0.213i$, $\tilde{n}=-1.027+0.104i$, and $m=n=0.1$ when $b=0$. The final upper bound is given by the smaller one of the two cases and the lower bound is given by the larger one, which are plotted in FIG.1. One can find that at the two endpoints (corresponding to $a=0$ and $b=0$ respectively), the two bounds are consistent with $|x|^{2}\Delta A^{2}+|y|^{2}\Delta B^{2}$. So an arbitrarily small deviation of the two endpoints will provide a
corresponding arbitrarily tight bound. This well shows the improvability of the bounds.

\section{The improvable bounds for uncertainty relation with multiple observables}

When more than two observables are taken into account, one will have to consider the uncertainty relation with multiple observables. Although uncertainty relation with two observables can automatically induce the corresponding uncertainty relations with more than two observables, one has to consider the different combinations of observables and find a maximal lower bound in order to reach a better tightness. Looking for proper combinations will become a complex job with the increasing number of observables. It is obvious that if our previous technique could be applied in this case, it will greatly reduce the complexity. This is because the bounds will be improvable once we find proper bounds using our technique. Next we present alternative lower and upper bounds for the sum of the variances of multiple observables.

\textbf{Theorem 2}: Suppose we measure multiple observables $\{A_{i},i=1,2,...,N\}$ on a normalized state $\left\vert \psi \right\rangle $, the stronger uncertainty relations are given by
\begin{equation}
\mathcal{\widetilde{B}}_{-}\leq \sum\limits_{i=1}^{N}\left\vert x_{i}\right\vert
^{2}\Delta ^{2}\left( A_{i}\right) \leq \mathcal{\widetilde{B}}_{+},
\label{theo2}
\end{equation}%
where
\begin{equation}
\mathcal{\widetilde{B}}_{\pm }=\frac{1}{N}\left[ \mathcal{\widetilde{B}}%
^{2}+\sum\limits_{k=1}^{N(N-1)/2}\left( \left\vert a_{k}\right\vert \mathcal{%
B}\left( m_{k},n_{k}\right) \mathcal{\pm }\left\vert b_{k}\right\vert
\mathcal{B}\left( \tilde{m}_{k},\tilde{n}_{k}\right) \right) ^{2}\right],
\label{bpmt}
\end{equation}%
and
\begin{eqnarray}
\mathcal{\widetilde{B}} &=&\Delta ^{2}\left(
\sum\limits_{j=1}^{N}x_{Rj}A_{j}\right) +\Delta ^{2}\left(
\sum\limits_{j=1}^{N}x_{Ij}A_{j}\right) +i\sum\limits_{j<l}\left(
x_{Rj}x_{Il}-x_{Ij}x_{Rl}\right) \left\langle \left[ A_{j},A_{l}\right]
\right\rangle ,  \label{btdefi}
\end{eqnarray}%
with $\mathcal{B}\left( \alpha ,\beta \right) $ defined as Theorem 1 and all the relevant free parameters confined by
\begin{eqnarray}
a_{k}m_{k}+b_{k}\tilde{m}_{k} &=&x_{i},  \label{pcons1} \\
a_{k}n_{k}+b_{k}\tilde{n}_{k} &=&-x_{j},  \label{pcons2}
\end{eqnarray}%
and
\begin{equation}
\left( \sum\limits_{i=1}^{N}\left\vert a_{k}\right\vert ^{2}\right) \left(
\sum\limits_{i=1}^{N}\left\vert b_{k}\right\vert ^{2}\right) \neq 0.
\label{limp}
\end{equation}

\emph{Proof.} We will prove this theorem parallel with the proof of theorem 1. Let's first show Eq. (\ref{theo2}) is valid. Define the unnormalized states as
\begin{equation}
|\psi _{i}\rangle =\left( A_{i}-\langle A_{i}\rangle \right) |\psi \rangle ,
\label{inid}
\end{equation}%
so $\Delta ^{2}\left( A_{i}\right) =\left\Vert |\psi _{i}\rangle \right\Vert^{2}$. Thus we can construct the following states
\begin{equation}
|\bar{\psi}\rangle =\sum\limits_{i=1}^{N}x_{i}|\psi _{i}\rangle.
\label{conis}
\end{equation}%
and
\begin{eqnarray}
\vert \tilde{\psi}_{k}\rangle &=&x_{i}|\psi _{i}\rangle
-x_{j}|\psi _{j}\rangle , 1 \leq i<j\leq N, k =1,\cdots ,\frac{N(N-1)}{2}.
\label{psi}
\end{eqnarray}%
Similarly to Eq. (\ref{recs}), we can reconstruct $\vert \tilde{\psi}_{k}\rangle $ as%
\begin{equation}
\vert \tilde{\psi}_{k}\rangle =a_{k}|\tilde{\psi}_{k1}\rangle
+b_{k}|\tilde{\psi}_{k2}\rangle,  \label{rrecs}
\end{equation}%
with
\begin{eqnarray}
\vert \tilde{\psi}_{k1}\rangle &=&m_{k}\vert \psi
_{i}\rangle +n_{k}\vert \psi _{j}\rangle ,  \label{rrs1} \\
\vert \tilde{\psi}_{k2}\rangle &=&\tilde{m}_{k}\vert \psi
_{i}\rangle +\tilde{n}_{k}\vert \psi _{j}\rangle .
\label{rrs2}
\end{eqnarray}%
Comparing Eq. (\ref{rrecs}) with Eq. (\ref{psi}), we can easily find that the complex parameters $a_{k}$, $b_{k}$, $m_{k}$, $n_{k}$, $\tilde{m}_{k}$, $\tilde{n}_{k}$ should satisfy%
\begin{eqnarray}
a_{k}m_{k}+b_{k}\tilde{m}_{k} &=&x_{i},  \label{lim11} \\
a_{k}n_{k}+b_{k}\tilde{n}_{k} &=&-x_{j}.  \label{lim22}
\end{eqnarray}%
According to $l_{2}$ norm of a state, one can obtain%
\begin{eqnarray}
\left\Vert |\bar{\psi}\rangle \right\Vert ^{2} &=&\left\Vert
\sum\limits_{i=1}^{N}x_{i}|\psi _{i}\rangle \right\Vert
^{2}=\sum\limits_{i=1}^{N}\left\vert x_{i}\right\vert ^{2}\left\Vert |\psi
_{i}\rangle \right\Vert ^{2}+\sum\limits_{i\neq j}x_{i}^{\ast }x_{j}\langle
\psi _{i}|\psi _{j}\rangle  \label{ex1}
\end{eqnarray}%
and for all $k,$%
\begin{eqnarray}
\Vert \vert \tilde{\psi}_{k}\rangle \Vert
^{2}=\Vert x_{i}\vert \psi _{i}\rangle \Vert
^{2}+\Vert x_{j}\vert \psi _{j}\rangle \Vert
^{2}-x_{i}^{\ast }x_{j}\langle \psi _{i}|\psi _{j}\rangle -x_{i}x_{j}^{\ast
}\langle \psi _{j}|\psi _{i}\rangle ,i\neq j  \label{ex2}
\end{eqnarray}%
Sum Eq. (\ref{psi}) and Eq. (\ref{ex2}), we will arrive at
\begin{equation}
\left\Vert |\bar{\psi}\rangle \right\Vert
^{2}+\sum\limits_{k=1}^{N(N-1)/2}\left\Vert \vert \tilde{\psi}%
_{k}\rangle \right\Vert ^{2}=N\sum\limits_{i=1}^{N}\left\Vert
x_{i}\left\vert \psi _{i}\right\rangle \right\Vert ^{2}.  \label{frfr}
\end{equation}%
Note that $\left\Vert |\bar{\psi}\rangle \right\Vert ^{2}$ can be expanded as%
\begin{eqnarray}
&&\left\Vert \vert \bar{\psi}\rangle \right\Vert ^{2}=\left\Vert
\sum\limits_{j=1}^{N}x_{j}|\psi _{j}\rangle \right\Vert ^{2}  \nonumber \\
&=&\left\Vert \sum\limits_{i=1}^{N}\left( x_{Rj}+ix_{Ij}\right) \left(
A_{j}-\langle A_{j}\rangle \right) \left\vert \psi \right\rangle
\right\Vert ^{2}  \nonumber \\
&=&\left\Vert \sum\limits_{j=1}^{N}x_{Rj}\left( A_{j}-\langle
A_{j}\rangle \right) \left\vert \psi \right\rangle
+\sum\limits_{j=1}^{N}ix_{Ij}\left( A_{j}-\langle A_{j}\rangle
\right) \left\vert \psi \right\rangle \right\Vert  \nonumber \\
&=&\Delta ^{2}\left( \sum\limits_{j=1}^{N}x_{Rj}A_{j}\right) +\Delta
^{2}\left( \sum\limits_{j=1}^{N}x_{Ij}A_{j}\right)+i\sum\limits_{j<l}\left(
x_{Rj}x_{Il}-x_{Ij}x_{Rl}\right) \left\langle \left[ A_{j},A_{l}\right]
\right\rangle  \nonumber \\
&=&\mathcal{\widetilde{B}}^{2},\label{btdef}
\end{eqnarray}%
where we use $x_{Rj}$ and $x_{Ij}$ to denote the real and imaginary parts of $x_{j}$. Meanwhile, we can also write all $\left\Vert \vert \tilde{\psi}_{k}\rangle \right\Vert ^{2}$ given in Eq. (\ref{rrecs}) as
\begin{eqnarray}
\left\Vert \vert \tilde{\psi}_{k}\rangle
\right\Vert^{2}&=&\left\Vert a_{k}|\tilde{\psi}_{k1}\rangle +b_{k}|\tilde{%
\psi}_{k2}\rangle \right\Vert ^{2} =\left\Vert a_{k}|\tilde{\psi}%
_{k1}\rangle \right\Vert ^{2}+\left\Vert b_{k}|\tilde{\psi}_{k2}\rangle
\right\Vert^{2}+2\mathrm{Re}\left( a_{k}^{\ast}b_{k}\langle \tilde{\psi}%
_{k1}|\tilde{\psi}_{k2}\rangle\right) .  \label{bdmm}
\end{eqnarray}%
Applying Cauchy-Schwartz inequality again, one has%
\begin{eqnarray}
\left( \left\Vert a_{k}|\tilde{\psi}_{k1}\rangle \right\Vert -\left\Vert
b_{k}|\tilde{\psi}_{k2}\rangle \right\Vert \right) ^{2} &\leq &\left\Vert
\vert \tilde{\psi}_{k}\rangle \right\Vert ^{2} \leq \left(
\left\Vert a_{k}|\tilde{\psi}_{k1}\rangle \right\Vert +\left\Vert b_{k}|%
\tilde{\psi}_{k2}\rangle \right\Vert \right) ^{2}.  \label{bdff}
\end{eqnarray}%
We can continue to expand $\left\Vert a_{k}|\tilde{\psi}_{k1}\rangle\right\Vert $ and $\left\Vert b_{k}|\tilde{\psi}_{k2}\rangle \right\Vert $ in terms of Eq. (\ref{rrs1}) and Eq. (\ref{rrs2}) as follows,
\begin{eqnarray}
\left\Vert a_{k}|\tilde{\psi}_{k1}\rangle \right\Vert &=&\left\vert
a_{k}\right\vert \mathcal{\widetilde{B}}\left( m_{k},n_{k}\right) ,  \label{bd1}
\\
\left\Vert b_{k}|\tilde{\psi}_{k2}\rangle \right\Vert &=&\left\vert
b_{k}\right\vert \mathcal{\widetilde{B}}\left( \tilde{m}_{k},\tilde{n}%
_{k}\right) .  \label{bd2}
\end{eqnarray}%
Insert Eq. (\ref{bd1}) and Eq. (\ref{bd2}) into Eq. (\ref{bdff}) and then substitute Eq. (\ref{bdff}) and Eq. (\ref{btdef}) into Eq. (\ref{frfr}), one will directly arrive at the result given in Eq. (\ref{theo2}).

Next we explain the two uncertainty bounds are also improvable and the lower bound is nontrivial. One can find that the inequalities come from the Cauchy-Schwartz inequality which is saturated when either all $a_k$ or all $b_k$ are simultaneously
vanish. With a single formula, one can express this condition just as Eq. (\ref{limp}) for nontrivial bounds. Thus, based on the continuous dependence on the parameters $a_k$ and $b_k$, one can draw the conclusion that both the bounds can be adjusted in the vicinities of $\left\vert a_k\right\vert$ and $\left\vert b_k\right\vert$ such that they infinitely approach the sum of variances so long as the sum does not vanish. So the bounds are improvable. Similarly, just because of the improvability, it implies that the lower bound will not be trivial since the vanishing sum of the variances is not a trivial condition for whether the considered state is the common eigenstate of the observables. The proof is finished.\hfill{}$\blacksquare$

\begin{figure}[tbp]
\centering
\includegraphics[width=0.5\columnwidth]{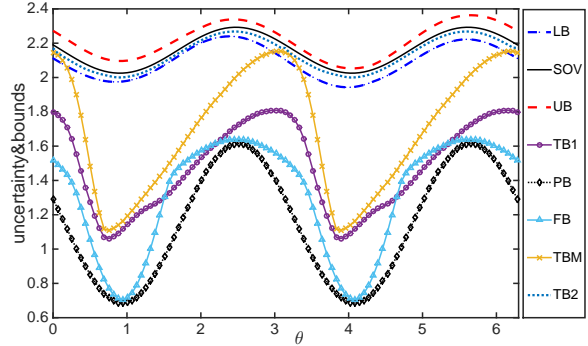}
\caption{(Color online.) The uncertainties of $A_{i}(i=1,2,3,4)$ with
various bounds vs. the parameter $\protect\theta$ of the quantum state. The
arbitrary tightness of our bounds is embodied by that the bounds can
infinitely approach to the exact SOV. This also demonstrates the powerful
technique we used here. }
\end{figure}

In order to demonstrate improvability of our bounds, we consider an example including four observables operated on a pure state. Simultaneously, we also compare our upper bound (UB) and lower bound (LB) in Eq. (\ref{theo2}) with the sum of the variances (SOV) and the bounds separately given in Ref. \cite{pati} (PB) and Ref. \cite{SRfei} (FB). Here the measured state is given by $\left\vert\psi\right\rangle=[\cos{\frac{\theta}{2}}e^{i\phi},\sin{\frac{\theta}{2}}]^T$ with a randomly generated phase $e^{i\phi}=0.6607 - 0.7507i$. The four observables are supposed to be $(A_{1},A_{2},A_{3})=(\sigma_{x},\sigma _{y},\sigma _{z})$ and
\begin{equation}
A_{4}=\left(
\begin{array}{cc}
0.8811 & 0.3876-0.2000i \\
0.3876 +0.2000i & 0.2403%
\end{array}%
\right),
\end{equation}
which is also randomly generated.

In principle, in order to get an ideally tight bound on the sum of the variances, one should find the global maximum of all the potential lower bounds. Here we only provide several lower bounds. First, directly applying our theorem 2, one can easily find the upper bound and the lower bound on the four observables. The concrete form can be obtained by substituting all the observables in Eq. (\ref{theo2}) and letting $x_{i}=1$ and $N=4$. In addition, one knows that the direct application of the uncertainty relations with two observables (In fact, we can also use the uncertainty relations with three observables, or other forms. How to group the observables could be complicated. The fundamental idea can also be found in our previous paper \cite{ZY}) can also induce the uncertainty with more observables. In our examples, this means that one should randomly select two observables from the set of $\{A_{1},A_{2},A_{3},A_{4}\}$. Here we only consider three of them for simplicity, that is,
\begin{equation}
\left\{
\begin{array}{c}
\{(A_{1},A_{2}),(A_{3},A_{4})\} \\
\{(A_{1},A_{3}),(A_{2},A_{4})\} \\
\{(A_{1},A_{4}),(A_{2},A_{3})\}%
\end{array}%
\right. .  \label{zuhe}
\end{equation}%
Based on theorem 1, we can immediately obtain the uncertainty relation for each pair of observables as $\Delta A_{i}^{2} +\Delta A_{j}^{2} \geq \mathcal{B}_{-}\left( A_{i},A_{j}\right) ,i\neq j$. Therefore, for each combination given by the $k$th line of Eq. (\ref{zuhe}), one always has an uncertainty relation with four observables as $\sum_{i=1}^{4}\Delta A_{i}^{2}\geq \mathcal{\widetilde{L}}_{k}$. So with all the three cases taken into account, one will arrive at
\begin{equation}
\sum_{i=1}^{4}\Delta A_{i}^{2} \geq \max \{\mathcal{\widetilde{L}}%
_{1},\mathcal{\widetilde{L}}_{2},\mathcal{\widetilde{L}}_{3}\}\geq \frac{1}{3}{%
\sum_{k=1}^{3}\mathcal{\widetilde{L}}_{k}.}  \label{twob}
\end{equation}%
For explicit, we would like to list the lower bounds separately presented by Ref. \cite{SRfei} and Ref. \cite{pati} . In Ref. \cite{SRfei} , the lower bound (FB) is given by%
\begin{equation}
\sum_{i=1}^{4}\Delta A_{i}^{2}\geq \frac{1}{2}\left\{ \sum_{1\leq i<j\leq
4}\Delta (A_{i}+A_{j})^{2}-\frac{1}{9}\left[ \sum_{1\leq i<j\leq 4}\Delta
(A_{i}+A_{j})\right] ^{2}\right\} .
\end{equation}
and in Ref. \cite{pati} , the lower bound (PB) is given by
\begin{equation}
\sum_{i=1}^{4}\Delta A_{i}^{2}\geq \frac{1}{6}\sum_{1\leq i<j\leq 4}\Delta
(A_{i}+A_{j})^{2}.  \label{patib}
\end{equation}

In FIG. 2, we plot all the mentioned bounds and the sum of the variances. For our upper bound (UB) and lower bound (LB), we use $k$ from $1$ to $6$ to order the different combinations of the observables $(A_1,A_2), (A_1,A_3),(A_1,A_4), (A_2,A_3), (A_2,A_4), (A_3,A_4)$. We set all $a_k=0.03,k=1,\cdots,6$, and randomly generate $b_k$, $m_k$, $n_k$ which are given in the current case by
\begin{equation}
\vec{b}= \left(%
\begin{array}{c}
0.1419 - 0.7572i \\
0.4064 - 0.1821i \\
0.5931 - 0.6196i \\
0.8094 - 0.7699i \\
0.4706 - 0.4211i \\
0.6741 - 0.4390i%
\end{array}%
\right), \vec{m} = \left(%
\begin{array}{c}
0.4888 + 0.8208i \\
0.2168 + 0.1228i \\
0.8329 + 0.6494i \\
0.5015 + 0.7366i \\
0.1027 + 0.7107i \\
0.3644 + 0.1053i%
\end{array}%
\right),\vec{n} = \left(%
\begin{array}{c}
0.9586 + 0.3085i \\
0.4237 + 0.1309i \\
0.5601 + 0.3435i \\
0.0988 + 0.6631i \\
0.4831 + 0.5162i \\
0.1536 + 0.7967i%
\end{array}%
\right).  \label{param2}
\end{equation}
In order to obtain two lower bounds given in Eq. (\ref{twob}), we also consider two cases. For simplicity, in the first case, we let all $a=0.5$, $b=5$, $m=2$, $n=1$ and one can obtain one lower bound. In the second case, we let all $a=0.5$, $b=5$, $\tilde{m}=1$, $\tilde{n}=1$ and one can obtain the other lower bound. The larger lower bound serves as the final lower bound. With this choice of the parameters, we use TBM to denote the lower bound by the maximization in Eq. (\ref{twob}) and use TB1 to denote the lower bound by the average in Eq. (\ref{twob}).

From FIG. 2, one can find that the lower bound (FB) given in Ref. \cite{SRfei} is strictly larger than the bound (PB) presented in Ref. \cite{pati}. However, our lower bound (LB) given in theorem 2 is much larger than both bounds (FB and PB) and quite close to $\sum_{i=1}^{4}\Delta A_{i}^2$ (SOV). It is interesting that the lower bound by two-to-two combination (TB1) is not always larger than FB, but the TBM with the same parameters in our examples is strictly larger than FB. In addition, one can also see that the lower bound TBM can sometimes go beyond the lower bound (LB) given in theorem 2. In this sense, it is difficult to roughly tell which method can provide a better lower bound. So in general case, one has to consider different methods and finally select the maximal lower bound as the final result. However, in this paper, one will find that our method to constructing the bounds is powerful. It will provide arbitrarily improvable bounds. To see this, we change the free parameters for the lower bound TB1 and now only set all $a=0.01$, $b=1$ and keep all other parameters unchanged. With these changed parameters, we use TB2 to represent this lower bound. It is obvious that the lower bound TB2 is the maximal one among all presented lower bounds. This just shows that the lower bound given by our method is arbitrarily tight. One should note that TB2 is larger than TBM because they have different free parameters. Once all the parameters are the same, TBM is always larger than TB2. This means that TBM can be also adjusted as close to SOV as possible. Of course, if we properly adjust the parameters, one can also let the lower bound LB given in theorem 2 close enough to SOV. Thus one sees that by applying our method, any one of the lower bounds with free parameters can serve as an arbitrarily improvable bound so long as the free parameters are properly adjusted. Finally, in FIG. 2 we also plot the upper bound (UB) given in theorem 2 as a demonstration of its improvability. Similarly, it can also be close enough to the SOV by adjusting the parameters.

\section{Discussions and Conclusion}

Before the end, we would like to emphasize that our method can be similarly used to construct the uncertainty relation for mixed states. It can be shown that the uncertainty relations for mixed states will have the same form as those for pure states. This can be done by replacing all the quantum states by the square roots of the corresponding density matrix. For example, in Eq. (\ref{s1}), $\left\vert\psi_1\right\rangle\rightarrow\sqrt{\rho_1}$ and $\left\vert\psi\right\rangle\rightarrow\sqrt{\rho}$. All the other states are similarly replaced. Following the same procedures as theorem 1 and theorem 2, one will arrive at the uncertainty relations with the same forms as Eq. (\ref{main}) and Eq. (\ref{theo2}). For pure states, if our method is understood based on the superposition principle of quantum states. For mixed states, one can understood as that any a matrix can be expanded by the `matrix basis'.

For the uncertainty relations with multiple observables, we have claimed that there is no unique way to constructing the bounds. In Theorem 2, we only give an alternative approach for the bounds. Besides the approach is relatively simple, the most important advantage is that the lower bound is always non-negative. One can also try other methods, for example, in Eq. (\ref{psi}) we can construct $|\tilde{\psi}_{i}\rangle =\sum_{j=1,j\neq i}^{N}a_{j}|\psi _{j}\rangle -a_{i}|\psi _{i}\rangle $. Correspondingly, one
will also get the bounds. But this method looks complex and especially it could lead to the negative lower bound. Of course, due to the improvability, one can also adjust the parameters such that the negative lower bound is avoided. In one word, no matter what form of the bounds one can construct, one will always obtain the improvable bounds so long as the essence of our method is employed.

To sum up, we have applied the superposition principle of quantum states to present the improvable lower and upper bounds for the uncertainty relation. It is shown that the bounds include free and controllable parameters so that they can be as tight as one expected. As a result, our lower bound only vanishes when the measured state is the common eigenstate of all the considered observables. In addition, due to the improvability, our method can greatly simplify the construction of bounds of uncertainty relation with multiple observables. That is, any one of the bound can always serve as a tight and nontrivial bound. Thus it seems that the uncertainty relations with multiple observables are not necessary, because it will be naturally covered by the uncertainty relation with only two observables. We also intuitively illustrate the improvability of our bounds by two examples separately for the uncertainty relations with two and four observables. Finally, we would like to say that many relevant and
interesting questions deserve us forthcoming efforts. Our bounds have many distinguished advantages, however, the cost is the introduction of many parameters. Therefore, what is the minimum number of the parameters for this kind of uncertainty relation with all our advantages? In addition, can our current technique be used to the other forms of uncertainty relations (principles)
including entropic uncertainty, noise-disturbance uncertainty? How can we consider the uncertainty relation with memory in the same manner?

\ack{This work was supported by the National Natural Science Foundation of China, under Grant No.11375036, the Xinghai Scholar Cultivation Plan and the Fundamental Research Funds for the Central Universities under Grants No. DUT15LK35 and No. DUT15TD47.}

\section*{Appendix}

\textbf{Lower bound of Eq. (\ref{patis}) in qubit system.-} At first, we would like to emphasize that the nontrivial bound doesn't only mean that a vanishing bound is given once the sum of the variances doesn't vanish. It will also be trivial if the lower bound is equal to the sum of variance, because in this case we doesn't need to look for \textit{bound}. Of course, this should be distinguished from the case when the state happens to be the eigenstate of one observable. With this in mind, let's study $\mathcal{L}_{1}$ and $\mathcal{L}_{2}$ in Eq. (\ref{patis}).

Since the qubit system is considered for Eq. (\ref{pati1}), there exists a unique $\vert \psi ^{\perp }\rangle $ orthogonal to $\left\vert \psi \right\rangle $ negleting a global phase. So we have $\vert \psi ^{\perp }\rangle\langle \psi ^{\perp}\vert =1-\left\vert \psi \right\rangle \left\langle \psi \right\vert $. Substituting this relation into Eq. (\ref{pati1}), one will have
\begin{eqnarray}
\mathcal{L}_{1} &=&\pm i\langle \lbrack X,Y]\rangle +|\langle \psi |X\pm
iY|\psi ^{\perp }\rangle |^{2}  \nonumber \\
&=&\pm i\langle \lbrack X,Y]\rangle +\langle \psi |\left( X\pm iY\right)
\left( X\mp iY\right) |\psi \rangle -\langle \psi |\left( X\pm iY\right) \left\vert \psi \right\rangle
\left\langle \psi \right\vert \left( X\mp iY\right) |\psi \rangle   \label{hengdeng}
\\
&=&\pm i\langle \lbrack X,Y]\rangle +\langle X^{2}\rangle
+\langle Y^{2}\rangle \mp i\left\langle [X,Y]\right\rangle  -\left\langle X\right\rangle ^{2}-\left\langle Y\right\rangle ^{2}
\nonumber \\
&=&\Delta X^{2} +\Delta  Y^{2} .  \label{yanzheng}
\end{eqnarray}%
In this sense, the maximum operation in Eq. (\ref{patis}) directly ignores $\mathcal{L}_{2}$ and becomes a trivial "bound". In this case, if one could give up $\mathcal{L}_{1}$ and turn to $\mathcal{L}_{2}$, this is also of no help, because $\mathcal{L}_{2}$ is trivial once $\left\vert \psi \right\rangle $ happens to be the eigenstate of $X+Y$, but it is not the
eigenstate of either $X$ or $Y$. So we think the bound given in Eq. (\ref{patis}) is trivial in qubit system. One could argue that the equality couldn't be a bad thing because it provides a unified form of Eq. (\ref{pati1}) for any qudit state. However, this is only a trivial substitution for qubit system. Considering our Eq. (\ref{hengdeng}) and Eq. (\ref{yanzheng}), one will find that they form an equality for any qudit state $\left\vert\psi\right\rangle$. In particular, all the terms in the left-hand-side of Eq. (\ref{hengdeng}) are the average values of observables such as $i\langle [X,Y]\rangle$, $\langle \psi |\left( X\pm iY\right)\left( X\mp iY\right) |\psi \rangle$, $\left\langle\psi\right\vert X\left\vert \psi\right\rangle$ and $\left\langle\psi\right\vert Y\left\vert\psi\right\rangle$. However, this is obviously an rewriting of Eq. (\ref{yanzheng}) and deviates the original purpose of uncertainty relation.

\end{document}